\documentclass[runningheads]{llncs}
\usepackage{numprint}
\usepackage{comment}
\usepackage{tablefootnote}
\usepackage{url}
\usepackage{tabularx}
\usepackage{todonotes}
\usepackage{booktabs}
\usepackage{subfigure}
\usepackage{cleveref} 
\usepackage{listings}
\usepackage{color}
\usepackage{multirow}
\usepackage{colortbl}
\usepackage{enumitem}
\usepackage[breaklinks]{hyperref}
\usepackage{txfonts}
\usepackage{fancybox}
\usepackage{hyperref}

\usepackage{xspace}
\usepackage[ruled,vlined, linesnumbered]{algorithm2e}

\definecolor{cycle3}{RGB}{77, 175, 74}
\newcommand{\win}{\cellcolor{cycle3!30}}

\newcommand{\emphb}[1]{\textbf{\textit{#1}}}
\newcommand{\defn}[1]{\emphb{#1}\quad}

\newcommand{\furl}[1]{\footnote{\scriptsize \url{#1}}}

\newcommand{\f}[1]{\footnote{\scriptsize #1}}
\newcounter{metric}

\makeatletter
\newcommand\newitem[1][]{\item[(#1)]\refstepcounter{metric}\def\@currentlabel{#1}}
\makeatother

\makeatletter
\newcommand\newMetricNr[1][]{#1\refstepcounter{metric}\def\@currentlabel{#1}}
\makeatother

\DeclareRobustCommand{\thetitle}{A Scalable Framework for Quality Assessment of RDF Datasets}

\begin{document}
\title{\thetitle}
\author{Gezim Sejdiu\inst{1}, Anisa Rula
\inst{1, 2}, Jens Lehmann\inst{1,3}  \and Hajira Jabeen\inst{1} }
\institute{
Smart Data Analytics, University of Bonn, Germany \\
\email{sejdiu@cs.uni-bonn.de},
\email{rula@cs.uni-bonn.de}, \email{jabeen@cs.uni-bonn.de}, \email{jens.lehmann@cs.uni-bonn.de} \and
University of Milano-Bicocca, Department of Computer Science, Systems and Communication (DISCo), Italy \\ \email{anisa.rula@disco.unimib.it} \and
Fraunhofer IAIS, Germany \\ \email{jens.lehmann@iais.fraunhofer.de}
}
\maketitle
\begin{tabular}{lcl}
\textbf{Resource type} & & Software Framework \\
\textbf{Website} & & \url{http://sansa-stack.net/distqualityassessment/}\\
\textbf{Permanent URL} & & \url{https://doi.org/10.6084/m9.figshare.7930139}
\end{tabular}

\begin{abstract}
Over the last years, Linked Data has grown continuously. 
Today, we count more than 10,000 datasets being available online following Linked Data standards. 
These standards allow data to be machine readable and inter-operable.  
Nevertheless, many applications, such as data integration, search, and interlinking, cannot take full advantage of Linked Data if it is of low quality.
There exist a few approaches for the quality assessment of Linked Data, but their performance degrades with the increase in data size and quickly grows beyond the capabilities of a single machine.
In this paper, we present DistQualityAssessment -- an open source 
implementation of quality assessment of large RDF datasets that can scale out to a cluster of machines.
This is the first distributed, in-memory approach for computing different quality metrics for large RDF datasets using Apache Spark. We also provide a quality assessment pattern that can be used to generate new scalable metrics that can be applied to big data.
The work presented here is integrated with the SANSA framework and has been applied to at least three use cases beyond the SANSA community.   
The results show that our approach is more generic, efficient, and scalable as compared to previously proposed approaches.

\end{abstract}

\section{Introduction}\label{sec:introduction}
Large amounts of data are being published openly to Linked Data by different data providers. 
A multitude of applications such as semantic search, query answering, and machine reading~\cite{rw2014} depend on these large-scale\furl{http://lodstats.aksw.org/} RDF datasets.  
The quality of underlying RDF data plays a fundamental role in large-scale data consuming applications. 
Measuring the quality of linked data spans a number of dimensions including but not limited to: accessibility, interlinking, performance, syntactic validity or completeness~\cite{zaveri2015quality}.
Each of these dimensions can be expressed through one or more quality metrics. 
Considering that each quality metric tries to capture a particular aspect of the underlying data, numerous metrics are usually provided against the given data that may or may not be processed simultaneously.

On the other hand, the limited number of existing techniques of quality assessment for RDF datasets are not adequate to assess data quality at large-scale and these approaches mostly fail to capture the increasing volume of big data. 
To date, a limited number of solutions have been conceived to offer quality assessment of RDF datasets \cite{Debattista0AC18,farber2018,beek2018,debattista2016luzzu}.
But, these methods can either be used on a small portion of large datasets \cite{farber2018} or narrow down to specific problems e.g., syntactic accuracy of literal values~\cite{beek2018}, or accessibility of resources~\cite{Mihindukulasooriya2016LDSA}.
In general, these existing efforts show severe deficiencies in terms of performance when data grows beyond the capabilities of a single machine.
This limits the applicability of existing solutions to medium-sized datasets only, in turn, paralyzing the role of applications in embracing the increasing volumes of the available datasets.

To deal with big data, tools like Apache Spark\furl{https://spark.apache.org/} have recently gained a lot of interest. 
Apache Spark provides scalability, resilience, and efficiency for dealing with large-scale data. Spark uses the concepts of Resilient Distributed Datasets (RDDs)~\cite{zaharia2012resilient} and performs operations like transformations and actions on this data in order to effectively deal with large-scale data. 

To handle large-scale RDF data, it is important to develop flexible and extensible methods that can assess the quality of data at scale. 
At the same time, due to the broadness and variety of quality assessment domain and resulting metrics, there is a strong need to provide a generic pattern
to characterize the quality assessment of RDF data in terms of scalability and applicability to big data.

In this paper, we borrow the concepts of data \textit{transformation} and \textit{action} from Spark and present a 
pattern for designing quality assessment metrics over large RDF datasets, which is inspired by design patterns.
In software engineering, design patterns are general and reusable solutions to common problems. 
Akin to design pattern, where each pattern acts like a blueprint that can be customized to solve a particular design problem, 
the introduced concept of Quality Assessment Pattern ($\mathcal{QAP}$) represents a generalized blueprint of scalable quality assessment metrics. 
In this way, the quality metrics designed following $\mathcal{QAP}$ can exhibit the ability to achieve scalability to large-scale data and work in a distributed manner.
In addition, we also provide an open source implementation and assessment of these quality metrics in Apache Spark following the proposed $\mathcal{QAP}$.

Our contributions can be summarized in the following points:
\begin{itemize}
 \item We present a Quality Assessment Pattern $\mathcal{QAP}$ to characterize scalable quality metrics.
 \item We provide DistQualityAssessment\furl{https://github.com/SANSA-Stack/SANSA-RDF/tree/develop/sansa-rdf-spark/src/main/scala/net/sansa_stack/rdf/spark/qualityassessment} -- a distributed (open source) implementation of quality metrics using Apache Spark.
 \item We perform an analysis of the complexity of the metric evaluation in the cluster.
 \item We evaluate our approach and demonstrate empirically its superiority over a previous centralized approach.
 \item We integrated the approach into the SANSA\furl{http://sansa-stack.net/} framework. 
 SANSA is actively maintained and uses the community ecosystem (mailing list, issues trackers, continues integration, web-site etc.).
 \item We briefly present three use cases where DistQualityAssessment has been used.
\end{itemize}

The paper is structured as follows:
Our approach for the computation of RDF dataset quality metrics is detailed in \autoref{sec:approach} and evaluated in \autoref{sec:evaluation}.
Related work on the computation of quality metrics for RDF datasets is discussed in \autoref{sec:related_work}.
Finally, we conclude and suggest planned extensions of our approach in \autoref{sec:conclusion}.

\section{Approach}\label{sec:approach}
In this section, we first introduce basic notions used in our approach, the formal definition of the proposed quality assessment pattern and then describe the workflow. 

\subsection{Quality Assessment Pattern}
Data quality is commonly conceived as a multi-dimensional construct~\cite{BatiniS16} with a popular notion of 'fitness for use' and can be measured along many dimensions $\mathcal{D}$ such as accuracy ($d_{accu} \in \mathcal{D}$), completeness ($d_{comp} \in \mathcal{D}$) and timeliness ($d_{tmls} \in \mathcal{D}$). 
The assessment of a quality dimensions $d$ is based on quality metrics $\mathcal{QM} = {m_1,m_2 …...m_k}$ where $m_i$ is a heuristic that is designed to fit a specific assessment dimension. 
The following definitions form the basis of $\mathcal{QAP}$.



\begin{definition}[Filter]
Let $\mathcal{F} = {f_1,f_2 …...f_l}$ be a set of filters where each filter $f_i$ sets a criteria for extracting predicates, objects, subjects, or their combination. 
A filter $f_i$ takes a set of RDF triples as input and returns a subgraph that satisfies the filtering criteria.
\end{definition}

\begin{definition}[Rule]
Let $\mathcal{R} = {r_1,r_2 …...r_j}$ be a set of rules where each rule $r_i$ sets a conditional criteria. A rule takes a subgraph as input and returns a new subgraph that satisfies the conditions posed by the rule $r_i$.
\end{definition}
\begin{definition}[Transformation]
\label{def:tr}
A \textit{transformation} $\tau:\mathcal{G} \rightarrow \mathcal{G'}$ is an operation that applies rules defined by $\mathcal{R}$ on the RDF graph $\mathcal{G}$ and returns an RDF subgraph $\mathcal{G'}$. 
A transformation $\tau$ can be a union $\cup$ or intersection $\cap$ of other transformations. 
\end{definition}
\begin{definition}[Action]
\label{def:ac}
An \textit{action} $ \alpha: \mathcal{G}\rightarrow \mathbb{R}$ is an operation that triggers the transformation of rules on the filtered RDF graph $\mathcal{G'}$ and generates a numerical value. 
Action $\alpha$ is the count of elements obtained after performing a $\tau$ operation. 
\end{definition}


\begin{definition}[Quality Assessment Pattern $\mathcal{QAP}$]
\label{def:QAP}
The Quality Assessment Pattern $\mathcal{QAP}$ is a reusable template to implement and design scalable quality metrics. 
The $\mathcal{QAP}$ is composed of \textit{transformations} and \textit{actions}. 
The output of a $\mathcal{QAP}$ is the outcome of an action returning a numeric value against the particular metric.
\end{definition}

$\mathcal{QAP}$ is inspired by Apache Spark operations and designed to fit different data quality metrics (for more details see \autoref{Table:QM}). 
Each data \textit{quality metric} can be defined following the $\mathcal{QAP}$. 
Any given data quality metric $m_i$ that is represented through the $\mathcal{QAP}$ using transformation $\tau$ and action $\alpha$ operations can be easily transformed into Spark code to achieve scalability.

\begin{table}[t]
\centering
\caption{Quality Assessment Pattern}\label{Table:QM}
\begin{tabular}{>{\scriptsize}l>{\scriptsize}l>{\scriptsize}l}
\toprule
    \verb|Quality Metric| &  \verb|:=| &  \verb|Action|   \textbar   \verb|(Action| $\mathcal{OP} $  \verb|Action)|\\
    $\mathcal{OP} $   &  \verb|:=| & $\mathcal{*}$ \textbar  $\mathcal{-}$ \textbar / \textbar $\mathcal{+}$ \\
 \verb|Action| &  \verb|:=| & \verb|Count(Transformation)| \\
 \verb|Transformation|  &  \verb|:=| &  \verb|Rule(Filter)| \textbar  \verb|(Transformation BOP Transformation)| \\
 \verb|Filter| &  \verb|:=| &  \verb|getPredicates|  $\sim ?p$ \textbar  \verb|getSubjects|  $ \sim ?s$ \textbar  \verb|getObjects| $\sim ?o$ \textbar  \verb|getDistinct(Filter)|\\
 && \textbar  \verb|Filter or Filter|  \textbar  \verb|Filter && Filter)|\\

 \verb|Rule| &  \verb|:=| &  \verb|isURI(Filter)| \textbar  \verb|isIRI(Filter)| \textbar  \verb|isInternal(Filter)| \textbar  \verb|isLiteral(Filter)|\\
&& \textbar  \verb|!isBroken(Filter)|  \textbar  \verb|hasPredicateP| \textbar   \verb|hasLicenceAssociated(Filter)| \\
&& \textbar  \verb|hasLicenceIndications(Filter)|  \textbar 
 \verb|isExternal(Filter)| \textbar  \verb|hasType((Filter)|\\
&&  \textbar \verb|isLabeled(Filter)| \\
 \verb|BOP| &  \verb|:=| & $\cap$ | $\cup$ \\

\end{tabular}
\end{table}

\autoref{tab:MetricRules} demonstrates a few selected quality metrics defined against proposed $\mathcal{QAP}$. 
As shown in \autoref{tab:MetricRules}, each quality metric can contain multiple rules, filters or actions. 
It is worth mentioning that action count(triples) returns the total number of triples in the given data. 
This can also be seen that the action can be an arithmetic combination of multiple actions i.e. ratio, sum etc. 
We illustrate our proposed approach on some metrics selected from~\cite{debattista2016luzzu,zaveri2015quality}. 
Given that the aim of this paper is to show the applicability of the proposed approach and comparison with existing methods, we have only selected those which are already provided out-of-box in Luzzu.

 \begin{table*}
    \centering
    \begin{tabular}{>{\scriptsize}l>{\scriptsize}l|>{\scriptsize}l>{\scriptsize}l|>{\scriptsize}l}
      \textbf{} & 
      \textbf{Metric} & 
      \multicolumn{2}{l|}{\textbf{\scriptsize Transformation $\tau$}} & 
      \textbf{Action $\alpha$} \\ 
        \hline  
        \newMetricNr[L1]\label{qm:L1} 
        & Detection of a & 
      \verb|r = hasLicenceAssociated(?p)| & & $\alpha$ \verb| = count(r)|  \\
       & 
     Machine Readable License & 
      & & $\alpha$ \verb| > 0 ? 1 : 0| \\
     \hline  
    \newMetricNr[L2]\label{qm:L2} 
    & Detection of a Human & 
    \verb|r = isURI(?s)| $ \cap $  \verb|hasLicenceIndications(?p)| $ \cap $  \verb|| & & $\alpha$ \verb| = count(r)| \\
    & Readable License & 
    \quad \quad \verb|isLiteral(?o)| $ \cap $  \verb|isLicenseStatement(?o)| & & $\alpha$ \verb| > 0 ? 1 : 0| \\
    \hline  
    \newMetricNr[I2]\label{qm:I2} 
    & Linkage Degree of Linked & 
      \verb|r_1 = isIRI(?s)| $\cap$  \verb|internal(?s)| $ \cap $ & 
      & $\alpha$\verb|_1 = count(r_3)| \\
    & External Data Providers  & 
     \quad \quad \quad \verb|isIRI(?o)| $ \cap $  \verb|external(?o)| & 
      & $\alpha$\verb|_2 = count(triples)|\\
    & & 
      \verb|r_2 = isIRI(?s)| $ \cap $  \verb|external(?s)| $ \cap $ & & $\alpha$\verb| = a_1/a_2| \\
    &  & 
     \quad \quad \quad \verb|isIRI(?o) | $ \cap $  \verb|internal(?o) | & 
      &  \\
    &  & 
      \verb|r_3 = r_1| $\cup$ \verb|r_2| & 
      & \\
    \hline  
    \newMetricNr[U1]\label{qm:U1} 
    & Detection of a Human & 
    \verb|r_1 = isURI(?s)| $ \cap $ \verb|isInternal(?s)| $ \cap $ & & $\alpha$\verb|_1 = count(r_1) +| \\
    & Readable Labels & 
     \quad \quad \quad  \verb|isLabeled(?p)| & & \quad \quad \quad \verb|count(r_2) +| \\
     &  & 
    \verb|r_2 = isInternal(?p)| $\cap$ \verb|isLabeled(?p)| & & \quad \quad \quad \verb|count(r_3)| \\
    & & 
    \verb|r_3 = isURI(?o)| $ \cap $ \verb|isInternal(?o)| $\cap$ & & $\alpha$\verb|_2 = count(triples)| \\
    & & \quad \quad \quad \verb|isLabeled(?p)| & & $\alpha$\verb|_1/| $\alpha$\verb|_2| \\
    \hline  
    \newMetricNr[RC1]\label{qm:RC1} 
    & Short URIs & 
      \verb|r_1 = isURI(?s)| $\cup$ \verb|isURI(?p)| $\cup$ \verb|isURI(?o)| & & $\alpha$\verb|_1 =count(r_2)| \\
    & & \verb|r_2 = resTooLong(?s, ?p, ?o)| & & $\alpha$\verb|_1/count(triples)| \\
    \hline  
    \newMetricNr[SV3]\label{qm:SV3} 
    & Identification of Literals & 
      \verb|r = isLiteral(?o)| $\cap$  \verb|getDatatype(?o)| $\cap$  &  & $\alpha$\verb| = count(r)| \\
    & with Malformed Datatypes & 
    \quad \quad \verb|isLexicalFormCompatibleWithDatatype(?o)| &  &   \\
     \hline  
    \newMetricNr[CN2]\label{qm:CN2} 
    & Extensional Conciseness & 
      \verb|r = isURI(?s)| $\cap$  \verb|isURI(?o)| &  & 
      $\alpha$\verb|_1 = count(r)| \\
      &  & &  & $\alpha$\verb|_2 = count(triples)| \\
      &  & &  &\verb|(|$\alpha$\verb|_2-| $\alpha$\verb|_1)/| $\alpha$\verb|_2| \\
      \end{tabular}
\caption{Definition of selected metrics following $\mathcal{QAP}$.}
\label{tab:MetricRules}
\end{table*}

\subsection{System Overview}

In this section, we give an overall description of the data model and the architecture of DistQualityAssessment.
We model and store RDF graphs $\mathcal{G}$ based on the basic building block of the Spark framework, RDDs. 
RDDs are in-memory collections of records that can be operated in parallel on a large distributed cluster.
RDDs provide an interface based on \emph{coarse-grained} transformations (e.g \emph{map}, \emph{filter} and \emph{reduce}): operations applied on an entire RDD. 
A \emph{map} function transforms each value from an input RDD into another value while applying $\tau$ rules.
A \emph{filter} transforms an input RDD to an output RDD, which contains only the elements that satisfy a given condition.
\emph{Reduce} aggregates the RDD elements using a specific function from $\tau$.

The computation of the set of quality metrics $\mathcal{QM}$ is performed using Spark as depicted in~\autoref{fig:DistQualityAssessmentSystem}.
Our approach consists of four steps: 

\begin{figure*}
\centering
\includegraphics[width=1.0\textwidth]{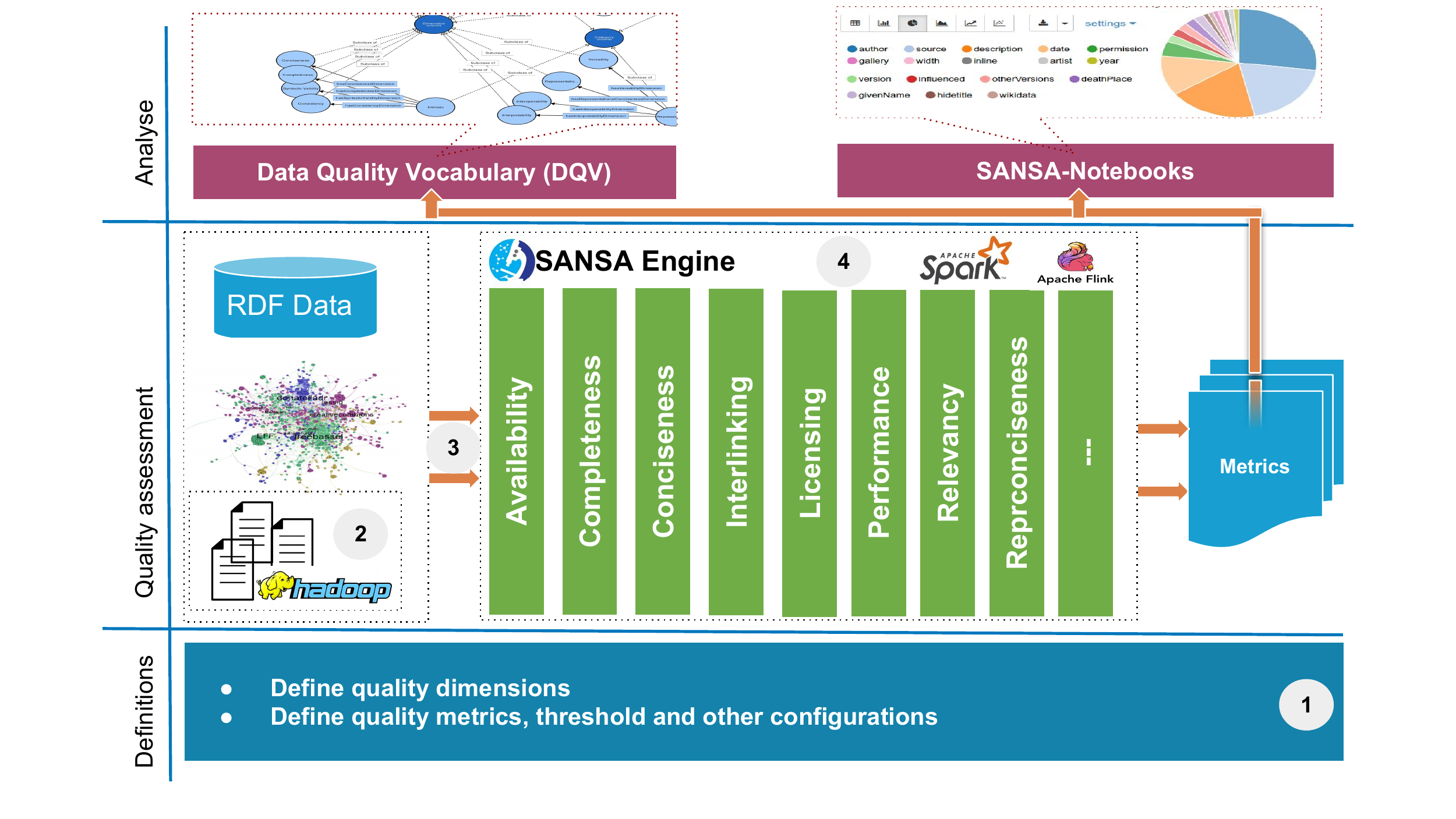}
\caption{Overview of distributed quality assessment's abstract architecture.}
\label{fig:DistQualityAssessmentSystem}
\end{figure*}


\paragraph{Defining quality metrics parameters (step 1)} The metric definitions are kept in a dedicated file which contains most of the configurations needed for the system to evaluate quality metrics and gather result sets.

\paragraph{Retrieving the RDF data (step 2)} RDF data first needs to be loaded into a large-scale storage that Spark can efficiently read from.
We use Hadoop Distributed File-System\furl{https://hadoop.apache.org/docs/r1.2.1/hdfs\_design.html} (HDFS).
HDFS is able to fit and stores any type of data in its Hadoop-native format and parallelizes them across a cluster while replicating them for fault tolerance.
In such a distributed environment, Spark automatically adopts different data locality strategies to perform computations as close to the needed data as possible in HDFS and thus avoids data transfer overhead.
 
\paragraph{Parsing and mapping RDF into the main dataset (step 3)} We first create a distributed dataset called \emph{main dataset} that represent the HDFS file as a collection of triples.
In Spark, this dataset is parsed and loaded into an RDD of triples having the format 
\emph{Triple$<$(s,p,o)$>$}.

\paragraph{Quality metric evaluation (step 4)} Considering the particular quality metric, Spark generates an execution plan
, which is composed of one or more $\tau$ transformations and $\alpha$ actions. The numerical output of the final action is the quality of the input RDF corresponding to the given metric.

\subsection{Implementation}
\label{subsection:implementation}

We have used the Scala\furl{https://www.scala-lang.org/} programming language API in Apache Spark to provide the distributed implementation of the proposed approach. 

The DistQualityAssessment (see \autoref{alg:DistQualityAssessment}) constructs the \emph{main dataset} (\autoref{line:rdf2rdd}) while reading RDF data (e.g. NTriples file or any other RDF serialization format) and converts it into an RDD of triples.
This latter undergoes the transformation operation of applying the filtering through rules in $R$ and producing a new \emph{filtered} RDD ($\mathcal{G'}$) (\autoref{line:filter}).
At the end, $\mathcal{G'}$ will serve as an input to the next step which applies a set of $\alpha$ actions (\autoref{line:action}).
The output of this step is the metric output represented as a numerical value (\autoref{line:action}). 
The result set of different quality metrics (\autoref{line:result}) can be further visualized and monitored using SANSA-Notebooks~\cite{iermilov-2017-sansa-iswc-demo}.\\
The user can also choose to extract the output in a machine-readable format (\autoref{line:dqvify}). 
We have used the data quality vocabulary\furl{https://www.w3.org/TR/vocab-dqv/} (DQV) to represent the quality metrics. 

\begin{algorithm}
\caption{Spark-based parallel quality assessment algorithm.}
\label{alg:DistQualityAssessment}
\SetKwInOut{Input}{input}\SetKwInOut{Output}{output}
\Input{$RDF$: an RDF dataset,
	   $param$: quality metrics parameters.
      }
\Output{$dqv$ description or $metric$ numerical value}
    $\textit{triples} = spark.\textbf{rdf}(lang)(input)$ \label{line:rdf2rdd} \\
    $\textit{triples}.persist()$ \label{line:cache}\\
    $dqv \leftarrow \emptyset$ \\
    \ForEach{$m \in param.getListOfMetrics$}{
        $triples \leftarrow triples.Tranform~\{~t =>$ \label{line:filter} \\
        $\quad \quad \textit{rule} \leftarrow m.Rule$ \\
        $\quad \quad t.apply(rule)~\}~$ \\
        $metric \leftarrow triples.apply(m.Action)$  \label{line:action}\\
        \If{m.hasDQVdescription}{
            $dqvify \leftarrow metric.dqvify()$ \label{line:dqvify}
        }
        $dqv.add(dqvify)$
    }
\Return{$(dqv, metric)$} \label{line:result}
\end{algorithm}

Furthermore, we also provide a Docker image of the system integrated within the BDE platform\furl{https://github.com/big-data-europe} - an open source Big Data processing platform allowing users to install numerous big data processing tools and frameworks and create working data flow applications.

The work done here (available under \textit{Apache License 2.0}) has been integrated into SANSA~\cite{lehmann-2017-sansa-iswc}, an open source\furl{https://github.com/SANSA-Stack} \emph{data flow processing engine} for scalable processing of large-scale RDF datasets. 
SANSA uses Spark 
offering fault-tolerant, highly available and scalable approaches to process massive sized datasets efficiently. SANSA provides the facilities for semantic data representation, querying, inference, and analytics at scale.
Being part of this integration, DistQualityAssessment can take advantage of having the same user community as well as infrastructure build via SANSA project.
Doing so, it can also ensure the sustainability of the tool given that SANSA is supported by several grants until at least 2021.

\paragraph{\textbf{Complexity Analysis}}
We deem that the overall time complexity of the distributed quality assessment evaluation is $O(n)$.
The performance of metrics computation depends on data shuffling (while filtering using rules in $R$) and data scanning. 
Our approach performs a direct mapping of any quality metric designed using $\mathcal{QAP}$ into a sequence of Spark-compliant Scala-commands, as a consequence, most of the operators used are a series of transformations like $map$, $filter$ and $reduce$.
The complexity of $map$ and $filter$ is considered to be linear with respect to the number of triples associated with it. 
The complexity of a metric then depends on the $\alpha$ operation that returns the count of the filtered output.
This later step works on the distributed RDD between $p$ nodes which imply that the complexity of each node then becomes $O(n/p)$, where $n$ is number of input triples.
Let be $O(\tau)$ a complexity of $\tau$, then the complexity of the metric will be $O(n/p*O(\tau))$.
This indicates that the runtime increases linearly when the size of an RDD increases and decreases linearly when more nodes $p$ are added to the cluster.

\section{Evaluation}\label{sec:evaluation}
The main aim of DistQualityAssessment is to serve massive large-scale real-life RDF datasets. 
We are interested in addressing the following additional questions.

\begin{itemize}
    \item \textbf{Flexibility}: How fast our approach processes different types of metrics?
    \item \textbf{Scalability}: How large are the RDF datasets that 
    DistQualityAssessment can scale to? 
    What is the system speedup w.r.t the number of nodes in a cluster mode?
    \item \textbf{Efficiency}: How well our approach performs compared with other state-of-the-art systems on real-world datasets?
\end{itemize}
In the following, we present our experimental setup including the datasets used. 
Thereafter, we give an overview of our results.

\subsection{Experimental Setup}

We chose two real-world and one synthetic datasets for our experiments:
 \begin{enumerate}
 \item \emph{DBpedia}~\cite{dbpedia-swj} (v 3.9) -- a cross domain dataset.
 DBpedia is a knowledge base with a large ontology.
 We build a set of 3 pipelines of increasing complexity: (i) $M_{DBpedia}^{en}$ ($\approx$ 813M triples); (ii) $M_{DBpedia}^{de}$ ($\approx$ 337M triples); (iii) $M_{DBpedia}^{fr}$ ($\approx$ 341M triples). 
DBpedia has been chosen because of its popularity in the Semantic Web community.
\item \emph{LinkedGeoData}~\cite{SLHA11} -- a spatial RDF knowledge base derived from OpenStreetMap.
\item \emph{Berlin SPARQL Benchmark (BSBM})~\cite{Bizer2009TheBS}  -- a synthetic dataset based on an e-commerce use case containing a set of products that are offered by different vendors and reviews posted by consumers about products.
The benchmark provides a data generator, which can be used to create sets of connected triples of any particular size.
 \end{enumerate}
Properties of the considered datasets are given in \autoref{tab:dataset_info}.

\begin{table*}
\centering
{\caption{Dataset summary information (nt format).}\label{tab:dataset_info}}
\begin{tabularx}{\textwidth}{Xcccccccc}	
\toprule
\multirow{2}{*}{$\longrightarrow$} & \multicolumn{1}{c}{} & \multicolumn{3}{c|}{DBpedia} & \multicolumn{4}{c}{BSBM} \\
\cline{3-9}  \rule{0pt}{10pt}
& LinkedGeoData & \scriptsize{en} & \scriptsize{de} & \scriptsize{fr}  & \scriptsize{2GB} &\scriptsize{20GB} &\scriptsize{200GB} &\\
\midrule
\scriptsize{\#nr. of triples}& \scriptsize{1,292,933,812} & \scriptsize{812,545,486} & \scriptsize{336,714,883} & \scriptsize{340,849,556} & \scriptsize{8,289,484} & \scriptsize{81,980,472} & \scriptsize{817,774,057} &  \\
\scriptsize{size (GB)} & \scriptsize{191.17} & \scriptsize{114.4} & \scriptsize{48.6} & \scriptsize{49.77} & \scriptsize{2} &\scriptsize{20} &\scriptsize{200} &\\
\bottomrule
\end{tabularx}
\end{table*}

We implemented DistQualityAssessment using Spark-2.4.0, Scala 2.11.11 and Java 8, and all the data were stored on the the HDFS cluster using Hadoop 2.8.0.
The experiments in local mode are all performed on a single instance of the cluster.
Specifically, we compare our approach with Luzzu~\cite{debattista2016luzzu} v4.0.0, a state-of-the-art quality assessment system\furl{https://github.com/Luzzu/Framework}.
All distributed experiments were carried out on a small cluster of 7 nodes (1 master, 6 workers): Intel(R) Xeon(R) CPU E5-2620 v4 @ 2.10GHz (32 Cores), 128 GB RAM, 12 TB SATA RAID-5.
The machines were connected via a Gigabit network.
All experiments have been executed three times and the average value is reported in the results.

\subsection{Results}

\begin{figure}
\centering
 \includegraphics[width=0.9\columnwidth]{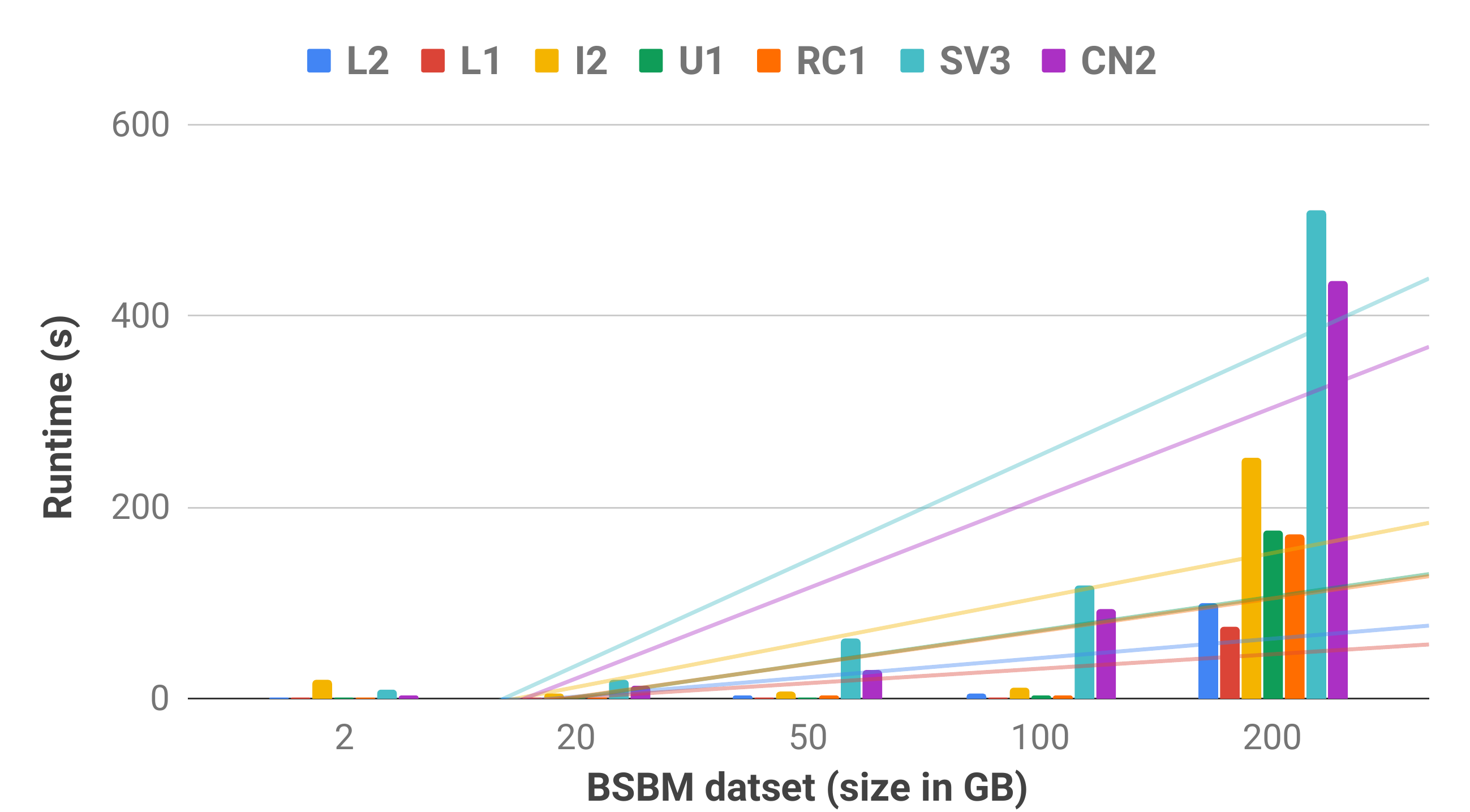}
    \caption{Sizeup performance evaluation.}
    \label{fig:sizeup-scalability}
\end{figure}

\begin{figure}
\centering
  \includegraphics[width=0.9\columnwidth]{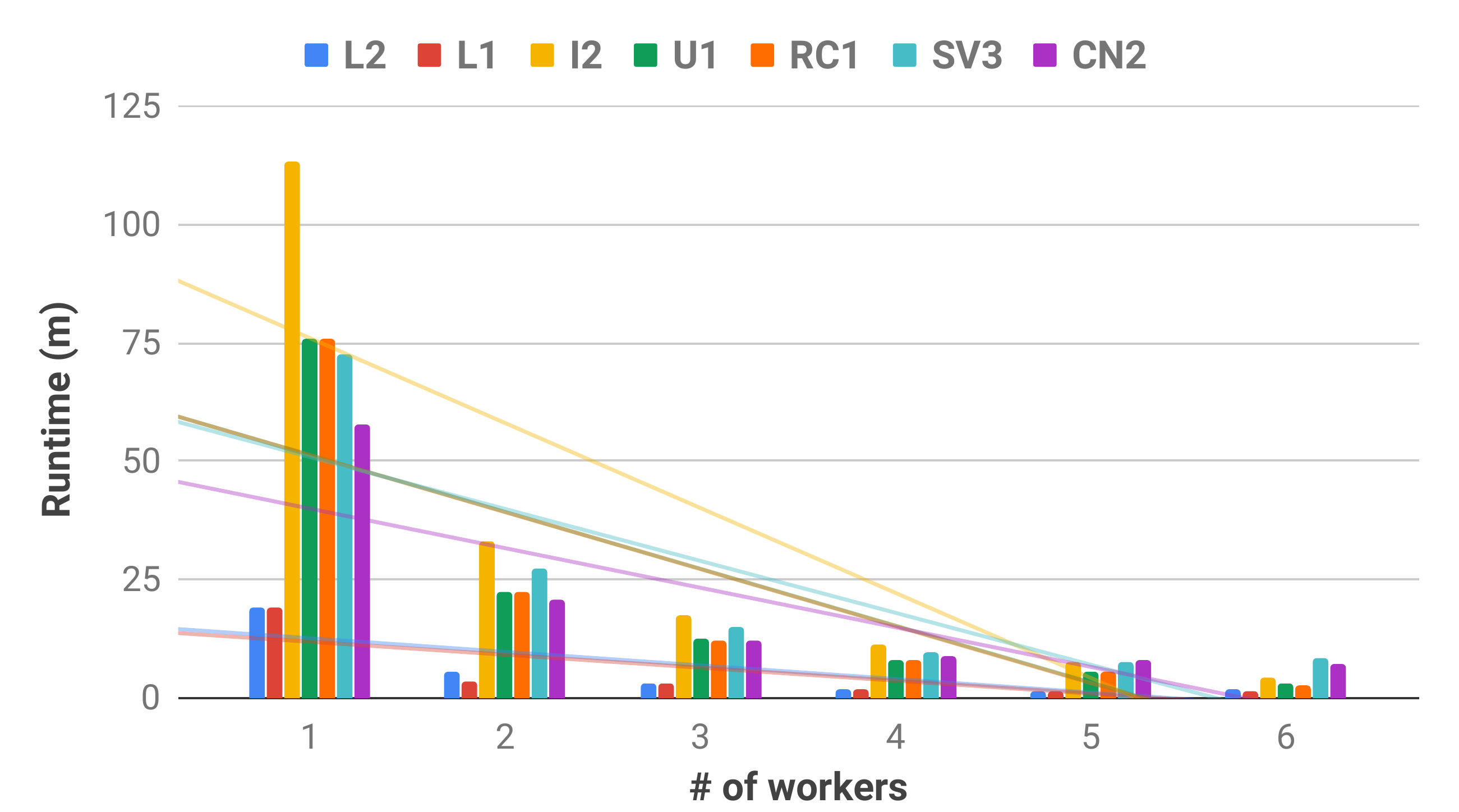}
    \caption{Node scalability performance evaluation.}
    \label{fig:node-scalability}
\end{figure}

We evaluate the proposed approach using the above datasets to compare it against Luzzu~\cite{debattista2016luzzu}.
We carry out two sets of experiments.
First, we evaluate the runtime of our distributed approach in contrast to Luzzu.
Second, we evaluate the horizontal scalability via increasing nodes in the cluster.
Results of the experiments are presented in \autoref{tbl:performance-evaluation}, \autoref{fig:sizeup-scalability} and
\autoref{fig:node-scalability}.
Based on the metric definition, some metrics make use of external access (e.g. Dereferenceability of Forward Links) which leads to a significant increase in Spark processing due to network latency. 
For the sake of the evaluation we have suspended such metrics.
As of that, we choose seven metrics (see \autoref{tab:MetricRules} for more details) where the level of difficulty vary from simple to complex according to combination of transformation/action operations involved.

\defn{Performance evaluation on large-scale RDF datasets}
\label{subsubsection:large_scale_datasets}
We started our experiments by evaluating the \textit{speedup} gained by adopting a distributed implementation of quality assessment metrics using our approach, and compare it against Luzzu.
We run the experiments on five datasets
($DBpedia_{en}$, $DBpedia_{de}$, $DBpedia_{fr}$, $LinkedGeoData$ and $BSBM_{200GB}$).
Local mode represent a single instance of the cluster without any tuning of Spark configuration and the cluster mode includes further tuning.
Luzzu was run in a local environment on a single machine with two strategies: (1) streaming the data for each metric separately, and (2) one stream/load -- all metrics evaluated just once. 

\begin{table*}
\centering
{\caption{Performance evaluation on large-scale RDF datasets.}\label{tbl:performance-evaluation}}
\begin{tabularx}{\textwidth}{Xcccccc}	
\toprule
\multicolumn{1}{l}{}& \multicolumn{5}{c}{\scriptsize{Runtime (m)} (\scriptsize{mean/std})} \\
\cline{2-6}
\rule{0pt}{8pt}
\multirow{2}{*}{$\longrightarrow$} & \multicolumn{2}{c|}{\scriptsize{\textbf{Luzzu}}} & \multicolumn{3}{c}{\scriptsize{\textbf{DistQualityAssessment}}} \\
\cline{2-6}  \rule{0pt}{10pt}
& \scriptsize{a) single} & \scriptsize{b) joint}  & \scriptsize{c) local} & \scriptsize{d) cluster} & \scriptsize{e) speedup ratio w.r.t} \\
& & & & & \scriptsize{Luzzu \textbar DistQualityAssessment$^{c)}$} \\
\midrule
\multirow{5}{*}{\rotatebox{90}{\textbf{Large-scale}}}
$LinkedGeoData$ & \scriptsize{Fail} & \scriptsize{Fail} & \scriptsize{446.9/63.34} & \win \scriptsize{7.79/0.54} & \win \scriptsize{n/a\textbar56.4x}\\
\hspace{0.2cm} $DBpedia_{en}$ & \scriptsize{Fail} & \scriptsize{Fail} & \scriptsize{274.31/38.17} & \win \scriptsize{1.99/0.04} & \win \scriptsize{n/a\textbar136.8x} \\
\hspace{0.2cm} $DBpedia_{de}$ & \scriptsize{Fail} & \scriptsize{Fail} & \scriptsize{161.4/24.18} & \win \scriptsize{0.46/0.04} & \win \scriptsize{n/a\textbar349.9x}\\
\hspace{0.2cm} $DBpedia_{fr}$ & \scriptsize{Fail} & \scriptsize{Fail} & \scriptsize{195.3/26.16} & \win \scriptsize{0.38/0.04} & \win \scriptsize{n/a\textbar512.9x}\\
\hspace{0.2cm} $BSBM_{200GB}$ & \scriptsize{Fail} & \scriptsize{Fail} & \scriptsize{454.46/78.04} & \win \scriptsize{7.27/0.64} & \win \scriptsize{n/a\textbar61.5x}\\
\midrule
\multirow{10}{*}{\rotatebox{90}{\textbf{Small to medium}}}
$BSBM_{0.01GB}$ & \scriptsize{2.64/0.02} & \scriptsize{2.65/0.01} & \win \scriptsize{0.04/0.0} & \scriptsize{0.42/0.04} & \win \scriptsize{65x\textbar (-0.9x)}\\
\hspace{0.2cm} $BSBM_{0.02GB}$ & \scriptsize{5.9/0.16} & \scriptsize{5.66/0.02} & \win \scriptsize{0.04/0.0} & \scriptsize{0.43/0.03} & \win \scriptsize{146.5x\textbar (-0.9x)}\\
\hspace{0.2cm} $BSBM_{0.05GB}$ & \scriptsize{16.38/0.44} & \scriptsize{15.39/0.21} & \win \scriptsize{0.05/0.0} & \scriptsize{0.46/0.02} & \win \scriptsize{326.6x\textbar (-0.9x)}\\
\hspace{0.2cm} $BSBM_{0.1GB}$ & \scriptsize{40.59/0.56} & \scriptsize{37.94/0.28} & \win \scriptsize{0.06/0.0} & \scriptsize{0.44/0.05} & \win \scriptsize{675.5x\textbar (-0.9x)}\\
\hspace{0.2cm} $BSBM_{0.2GB}$ & \scriptsize{101.8/0.72} & \scriptsize{101.78/0.64} & \win \scriptsize{0.07/0.0} & \scriptsize{0.4/0.03} & \win \scriptsize{1453.3\textbar (-0.8x)}\\
\hspace{0.2cm} $BSBM_{0.5GB}$ & \scriptsize{459.19/18.72} & \scriptsize{468.64/21.7} & \win \scriptsize{0.15/0.01} & \scriptsize{0.48/0.03} & \win \scriptsize{3060.3x\textbar (-0.7x)}\\
\hspace{0.2cm} $BSBM_{1GB}$ & \scriptsize{1454.16/10.55} & \scriptsize{1532.95/51.6} & \win \scriptsize{0.4/0.02} & \scriptsize{0.56/0.02} & \win \scriptsize{3634.4x\textbar (-0.3x)}\\
\hspace{0.2cm} $BSBM_{2GB}$ & \scriptsize{Timeout} & \scriptsize{Timeout} & \scriptsize{3.19/0.16} & \win \scriptsize{0.62/0.04} & \win \scriptsize{n/a\textbar 4.1x}\\
\hspace{0.2cm} $BSBM_{10GB}$ & \scriptsize{Timeout} & \scriptsize{Timeout} & \scriptsize{29.44/0.14} & \win \scriptsize{0.52/0.01} & \win \scriptsize{n/a\textbar 55.6x}\\
\hspace{0.2cm} $BSBM_{20GB}$ & \scriptsize{Fail} & \scriptsize{Fail} & \scriptsize{34.32/9.22} & \win \scriptsize{0.75/0.29} & \win \scriptsize{n/a\textbar 44.8x}\\
\bottomrule
\end{tabularx}
\end{table*}

\autoref{tbl:performance-evaluation} shows the performance of two approaches applied to five datasets.
In \autoref{tbl:performance-evaluation} we indicate "Timeout" whenever the process did not complete within a certain amount of time\f{We set the timeout delay to 24 hours of the quality assessment evaluation stage.} and "Fail" when the system crashed before this timeout delay.
Column Luzzu$^{a)}$ represents the performance of Luzzu on bulk load -- considering each metric as a sequence of the execution, on the other hand, the column Luzzu$^{b)}$ reports on the performance of Luzzu using a joint load by evaluating each metric using one load.
The last columns reports on the performance of DistQualityAssessment run on a local mode $c)$, cluster mode $d)$ and speedup ratio of our approach compared to Luzzu$^{b)}$ ($d)/b)-1$) and itself evaluated on local mode ($d)/c)-1$) is reported on the column $e)$.
We observe that the execution of our approach finishes with all the datasets whereas this is not the case with Luzzu which either timeout or fail at some point.


Unfortunately, Luzzu was not capable of evaluating the metrics over large-scale RDF datasets from \autoref{tbl:performance-evaluation} (part one). 
For that reason we run yet another set of experiments on very small datasets which Luzzu was able to handle. 
Second part of the \autoref{tbl:performance-evaluation} shows a performance evaluation of our approach compared with Luzzu on very small RDF datasets.
In some cases (e.g. \ref{qm:RC1}, \ref{qm:SV3}) for a very small dataset 
Luzzu performs better than our approach with a small margin of runtime in the local mode.
It is due to the fact that in the streaming mode, when Luzzu$^{a)}$ finds the first statement which fulfills the condition (e.g.finding the shortest URIs), it stops the evaluation and return the results.
On the contrary, our approach evaluates the metrics over the whole dataset exploiting the fault-tolerance and resilient features build in Spark.
In other cases Luzzu suffers from significant slowdowns, which are several orders of magnitude slower.
Therefore, its average runtime over all metrics is worst as compared to our approach. 
It is important to note that our approach on these very small datasets degrades while running on the cluster mode.
This is because of the network overhead while shuffling the data, but it outperforms Luzzu$^{a),b)}$ when considering ''average runtime'' over all the metrics (even for very small datasets).

Findings shown in \autoref{tbl:performance-evaluation} depict that our approach starts outperforming when the size of the dataset grows (e.g. $BSBM_{2GB}$).
The runtime in the cluster mode stays constant when the size of the data fits into the main memory of the cluster.
On other hand, Luzzu is not able to evaluate the metrics when the size of data starts increasing, the time taken lasts beyond the delay we set for small datasets. 
Because of the large differences, we have used a logarithmic scale to better visualize these results.

\defn{Scalability performance analysis}
\label{subsubsection:scalability_performance}
In this experiment we evaluate the efficiency of our approach. \autoref{fig:sizeup-scalability} and \autoref{fig:node-scalability} illustrates the results of the comparative efficiency analysis.


\textit{Data scalability} 
To measure the performance of \textit{size-up} scalability of our approach, we run experiments on five different sizes.
We fix the number of nodes to 6 and grow the size of datasets to measure whether DistQualityAssessment can deal with larger datasets.
For this set of experiments we consider BSBM benchmark tool to generate syntethic datasets of different sizes, since the real-world dataset are considered to be unique in their size and attributes.

We start by generating a dataset of 2GB.
Then, we iteratively increase the size of datasets.
On each dataset, we run our approach and the runtime is reported on \autoref{fig:sizeup-scalability}.
The $x$-axis shows the size of BSBM dataset with an increasing order of 10x magnitude.

By comparing the runtime (see \autoref{fig:sizeup-scalability}), we note that the execution time increases linearly and is near-constant when the size of the dataset increases.
As expected, it stays near-constant as long as the data fits in memory.
This demonstrates one of the advantages of utilizing the in-memory approach for performing the quality assessment computation.
The overall time spent in data read/write and network communication found in disk-based approaches is saved.
However, when the data overflows the memory, and it is spilled to disk, the performance degrades.
These results show the scalability of our algorithm in the context of size-up.

\textit{Node scalability} In order to measure node scalability, we vary the number of the workers on our cluster.
The number of workers have varied from 1, 2, 3, 4 and 5 to 6.

\autoref{fig:node-scalability} shows the speedup for $BSBM_{200GB}$ with the various number of worker nodes.
We can see that as the number of workers increases, the execution time cost-decrease is almost linear.
The execution time decreases about 14 times (from 433.31 minutes down to 28.8 minutes) as cluster nodes increase from one to six worker nodes.
The results shown here imply that our approach can achieve near linear scalability in performance in the context of speedup.

\begin{figure}
 \begin{minipage}{.5\textwidth}
\includegraphics[width=1\columnwidth]{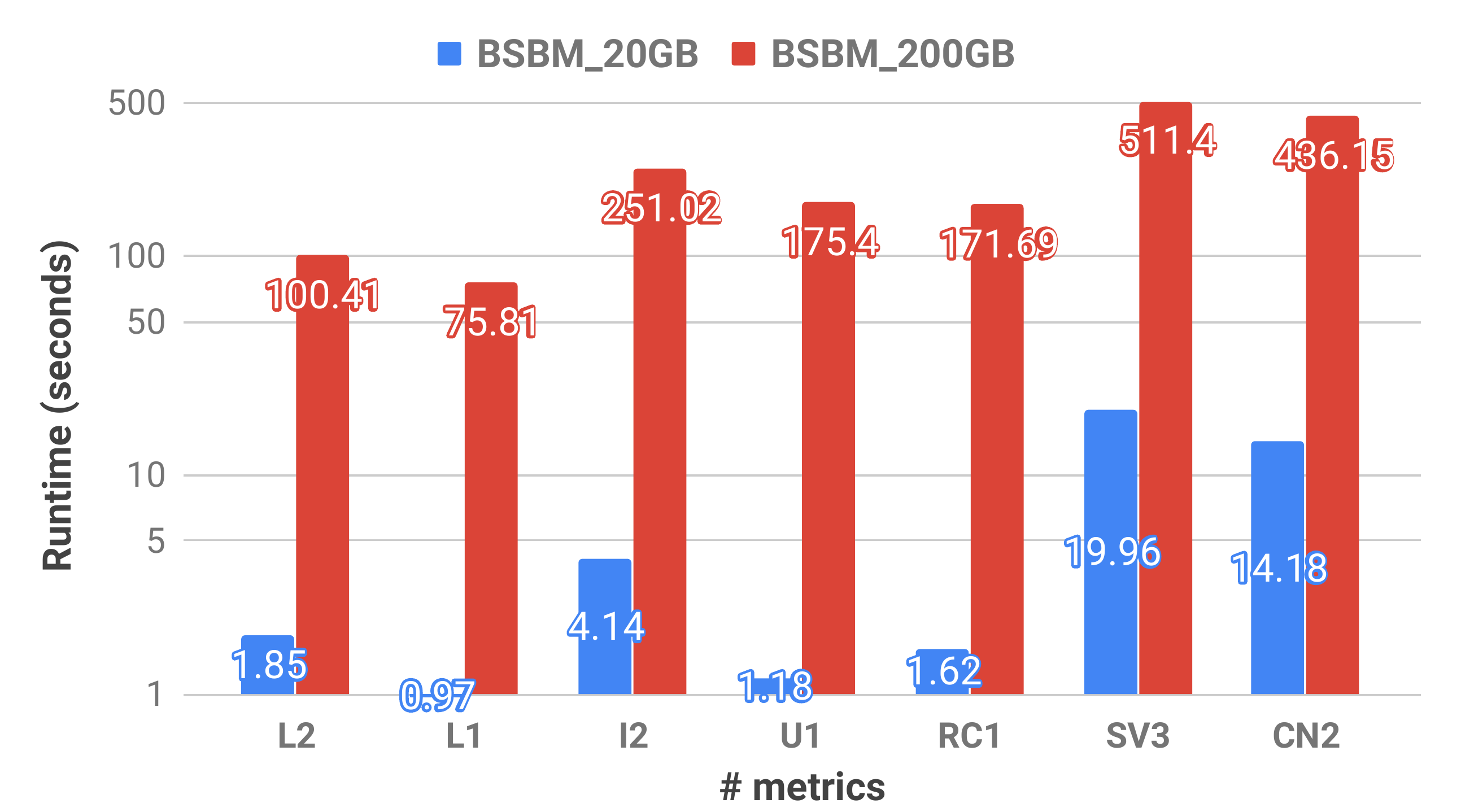}
\caption{Overall analysis by metric in the cluster mode (log scale).}
	\label{fig:overall-analysis}
 \end{minipage}
 \begin{minipage}{.5\textwidth}
\includegraphics[width=1.0\columnwidth]{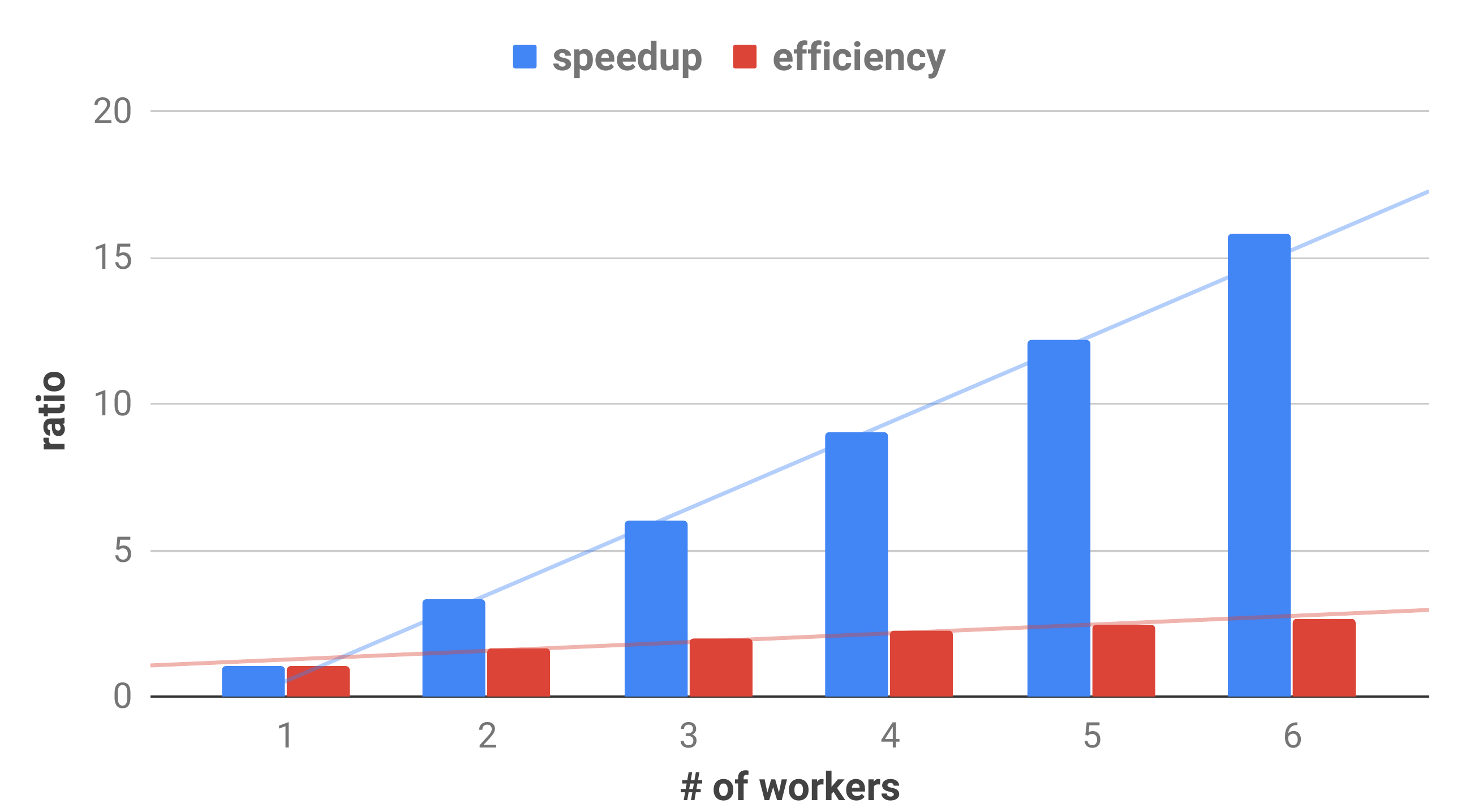}
    \caption{Effectiveness of DistQualityAssessment.}
    \label{fig:effectiveness}
 \end{minipage}
\end{figure}

Furthermore, we conduct the effectiveness evaluation of our approach.
Speedup $S$ is an important metric to evaluate a parallel algorithm.
It is defined as a ratio $S=T_s/T_n$, where $T_s$ represents the execution time of the algorithm run on a single node and $T_n$ represents the execution time required for the same algorithm on $n$ nodes with the same configuration and resources.
Efficiency is defined as a ratio $E = S/n =T_s/n T_n$ which measures the processing power being used, in our case the speedup per node.
The speedup and efficiency curves of DistQualityAssessment are shown in \autoref{fig:effectiveness}.
The trend shows that it achieves almost linearly speedup and even super linear in some cases.
The upper curve in the \autoref{fig:effectiveness} indicates super linear speedup. 
The speedup grows faster than the number of worker nodes.
This is due to the computation task for the metric being computationally intensive, and the data does not fit in the cache when executed on a single node. 
But it fits into the caches of several machines when the workload is divided amongst the cluster for parallel evaluation.
While using Spark, the super linear speedup is an outcome of the improved complexity and runtime, in addition to efficient memory management behavior of the parallel execution environment.

\defn{Correctness of metrics}
In order to test the correctness of implemented metrics, we assess the numerical values for metrics like \ref{qm:L1}, \ref{qm:L2}, and \ref{qm:RC1} on very small datasets and the results are found correct w.r.t Luzzu. 
For metrics like \ref{qm:I2} and \ref{qm:CN2}, Luzzu uses approximate values for faster performance, and that is not the same as getting the exact number as in the case of our implementation.

\defn{Overall analysis by metrics}
We analyze the overall run-time of the metric evaluation.
\autoref{fig:overall-analysis} reports on the run-time of each metric considered in this paper (see \autoref{tab:MetricRules}) on both $BSBM_{20GB}$ and $BSBM_{200GB}$ datasets.

DistQualityAssessment implements predefined quality assessment metrics from \cite{zaveri2015quality}. 
We have implemented these metrics in a distributed manner such that most of them have a run-time complexity of $O(n)$ where $n$ is the number of input triples.
The overall performance of analysis for BSBM dataset with two instances is shown in \autoref{fig:overall-analysis}.
The results obtained show that the execution is sometimes a little longer when there is a shuffling involved in the cluster compared to when data is processed without movement e.g.~Metric \ref{qm:L2} and \ref{qm:L1}.
Metric~\ref{qm:SV3} and \ref{qm:CN2} are the most expensive ones in terms of runtime.
This is due to the extra overhead caused by extracting the literals for objects, and checking the lexical form of its datatype. 

Overall, the evaluation study carried out in this paper demonstrates that distributed computation of  different quality measures is scalable and the execution ends in reasonable time given the large volume of data.

\section{Use Cases} \label{sec:use_cases}
The proposed quality assessment tool is being used in many use cases. These includes the projects QROWD, SLIPO, and an industrial application by Alethio\furl{https://goo.gl/mJTkPp}.

\defn{QROWD -- Crowdsourcing Streaming Big Data Quality Assessment Use Case}
QROWD\furl{http://qrowd-project.eu/} is a cross-sectoral streaming Big Data integration project including geographic, transport, meteorological, cross domain and news data, aiming to capitalize on hybrid Big Data integration and analytics methods.
One of the major challenges faced in QROWD, is to investigate options for effective and scalable data quality assessment on integrated (RDF) datasets using their crowdsourcing platform.
In order to perform this task efficiently and effectively, QROWD uses DistQualityAssessment as an underlying quality assessment framework.

\defn{Blockchain -- Alethio Use Case} 
Alethio\furl{https://aleth.io/} has build an Ethereum analytics platform that strives to provide transparency over the transaction pool of the whole Ethereum ecosystem. 
Their 18 billion triple data set\furl{https://medium.com/alethio/ethereum-linked-data-b72e6283812f} contains large scale blockchain transaction data modelled as RDF according to the structure of the Ethereum ontology\furl{https://github.com/ConsenSys/EthOn}.
Alethio is using SANSA in general and DistQualityAssesment in particular, for performing large-scale batch quality checks, e.g.~analysing the quality while merging new data, computing attack pattern frequencies and fraud detection. 
Alethio uses DistQualityAssesment on a cluster of 100 worker nodes to assess the quality of their $\approx$7 TB of data.

\defn{SLIPO -- Scalable Integration and Quality Assured fusion of Big POI data}
SLIPO\furl{http://slipo.eu/} is a project which leverages semantic web technologies for scalable and quality assured integration of large Point of Interest (POI) datasets.
One of the key features of the project is the fusion process.
SLIPO-fusion receives two different RDF datasets containing POIs and their properties, as well as a set of links between POI entities of the two datasets.
SLIPO is using DistQualityAssessment to assess the quality of both input datasets. The SLIPO-fusion produces a third, final dataset, containing consolidated descriptions of the linked POIs.
This process is often data and processing intensive, therefore, it requires a scalable mechanism for data quality check.
SLIPO uses DistQualityAssessment for fusion validation and quality statistics/assessment to facilitate and assure the quality of the fusion process.



\section{Related Work} \label{sec:related_work}
Even though quality assessment of big datasets is an important research area, it is still largely under-explored. 
There have been a few works discussing the challenges and issues of big data quality~\cite{becker2015big,RaoG015,cai2015challenges}. 
Only recently, a few of them have started to address the problem from a practical point of view~\cite{debattista2016luzzu}, which is the focus of our work as stated in \autoref{sec:introduction}. 
In the following, we divide the section between conceptual and practical approaches proposed in the state of the art for big data quality assessment.
In~\cite{CatarciSCD17} the authors propose a big data processing pipeline and a big data quality pipeline. 
For each of the phases of the processing pipeline they discuss the corresponding phase of the big data quality pipeline.
Relevant quality dimensions such as accuracy, consistency and completeness are discussed for the quality assessment of RDF datasets as part of an integration scenario.
Given that the quality dimensions and metrics have somehow evolved from relational to Linked Data,
it is relevant to understand the evolution of quality dimensions according to the differences between the structural characteristics of the two data models~\cite{BatiniRSV15}. 
This allows to manage the huge variability of methods and techniques needed to manage data quality and understand which are the quality dimensions that prevail when assessing large-scale RDF datasets. 

Most of the existing approaches can be applied to small/medium scale datasets and do not horizontally scale~\cite{debattista2016luzzu,KontokostasWAHLCZ14}. 
The work in~\cite{KontokostasWAHLCZ14} presents a methodology for assessing the quality of Linked Data based on a test case generation analogy used for software testing. 
The idea of this approach is to generate templates of the SPARQL queries (i.e., quality test case patterns) and then instantiate them by using the vocabulary or schema information, thus producing quality test case queries. 
Luzzu~\cite{debattista2016luzzu} is similar in spirit with our approach in that its objective is to provide a framework for quality assessment. 
In contrast to our approach, where data is distributed and also the evaluation of metrics is distributed, Luzzu does not provide any large-scale processing of the data. 
It only uses Spark streaming for loading the data which is not part of the core framework. 
Another approach proposed for assessing the quality of large-scale medical data implements Hadoop Map/Reduce~\cite{BonnerMKBTMCA15}. 
It takes advantage of query optimization and join strategies which are tailored to the structure of the data and the SPARQL queries for that particular dataset. In addition, this work, differently from our approach, does not assess any data quality metric defined in~\cite{zaveri2015quality}.
The work in~\cite{BenbernouO17} propose a reasoning approach to derive inconsistency rules and implements a Spark-based implementation of the inference algorithm for capturing and cleaning inconsistencies in RDF datasets. 
The inference generally incurs higher complexity. Our approach is designed for scalability, and we also use Spark-based implementation for capturing inconsistencies in the data. While the approach in~\cite{BenbernouO17} needs manual definitions of the inconsistency rules, our approach runs automatically, not only for consistency metrics but also for other quality metrics. 
In addition, we test the performance of our approach on large-scale RDF datasets while their approach is not experimentally evaluated. 
LD-Sniffer~\cite{Mihindukulasooriya2016LDSA}, is a tool for assessing the accessibility of Linked Data resources according to the metrics defined in the Linked Data Quality Model. 
The limitation of this tool, besides that it is a centralized version, is that it does not provide most of the quality assessment metrics defined in~\cite{zaveri2015quality}. 
In addition to above, there is a lack of unified structure to propose and develop new quality metrics that are scalable and less computationally expensive.

Based on the identified limitations of these aforementioned approaches, we have introduced DistQualityAssessment which bases its computation and evaluations mainly in-memory.
As a result the computation of the quality metrics show a high performance for large-scale datasets.

\section{Conclusions and Future Work}\label{sec:conclusion}
The data quality assessment becomes challenging with the increasing sizes of data.
Many existing tools mostly contain a customized data quality functionality to detect and analyze data quality issues within their own domain. 
However, this process is both data-intensive and computing-intensive and it is a challenge to develop fast and efficient algorithms that can handle large scale RDF datasets.

In this paper, we have introduced DistQualityAssessment, a novel approach for distributed in-memory evaluation of RDF quality assessment metrics implemented on top of the Spark framework.
The presented approach offers generic features to solve common data quality checks.
As a consequence, this can enable further applications to build trusted data utilities. 

We have demonstrated empirically that our approach improves upon previous centralized approach that we have compared against.
The benefit of using Spark is that its core concepts (RDDs) are designed to scale horizontally. Users can adapt the cluster sizes corresponding to the data sizes, by dropping when it is not needed and adding more when there is a need for it.

Although we have achieved reasonable results in terms of scalability, we plan to further improve time efficiency by applying intelligent partitioning strategies and persist the data to an even higher extent in memory and perform dependency analysis in order to evaluate multiple metrics simultaneously. We also plan to explore near real-time interactive quality assessment of large-scale RDF data using Spark Streaming.
Finally, in the future we intend to develop a declarative plugin for the current work using Quality Metric Language (QML)~\cite{debattista2016luzzu}, which gives users the ability to express, customize and enhance quality metrics.


\section*{Acknowledgment}
This work was partly supported by the EU Horizon2020 projects BigDataOcean (GA no.~732310), Boost4.0 (GA no.~780732), QROWD (GA no.~723088) and CLEOPATRA (GA no.~812997).

\bibliographystyle{abbrv}
\bibliography{bibliography}


\end{document}